# Density-functional study of lanthanide monoxides under high-pressure: Pressure-induced B1-B2 transition

S. Ferrari[1,2] and D. Errandonea[3,*]

[1] Departamento Física Experimental, CNEA, Centro Atómico Constituyentes, Av. Gral. Paz 1499, 1650 San Martín, Argentina

[2] Consejo Nacional de Ciencia y Tecnología, Argentina.

[2] Departamento de Física Aplicada-ICMUV, MALTA Consolider Team, Universidad de Valencia, Dr Moliner 50, Burjassot, 46100 Valencia, Spain

**Abstract:** Using density-functional theory we have studied the influence of pressure in crystal structure of lanthanide monoxides considering the fifteen elements of the lanthanide series, from La to Lu. Calculations have been performed using both the general-gradient (GGA) approximation and the local-density (LDA) approximation for the ambient-pressure B1 (NaCl-type) structure. By a systematic comparison with existent experimental data, we have found that the first method agrees better with them. In addition, considering other cubic structures previously reported for lanthanide monoxides, as B2 (CsCl-type) and B3 (ZnS-type), we have explored the possibility of pressure-induced phase transitions. Based on the better accuracy of GGA to describe the B1 phase, we have exclusively used GGA for the high-pressure study. We have found for the fifteen studied compounds that at ambient pressure the B1 structure is the one with the lowest enthalpy, being therefore the most stable one. We have also determined that at elevated pressures all the studied compounds undergo a phase transition to the B2 phase. We have finally established the relationship between pressure and the volume of the unit cell, along with the associated isothermal equation of state.

*Corresponding author, email: daniel.errandonea@uv.es

# 1. Introduction

Divalent lanthanide monoxides are known to exist from the seventies or even before [1,2,3]. However, due to the challenges faced to synthesize them and their weak chemical stability, their solid forms have not been studied until recently [4,5,6]. Despite these efforts, little is known about their crystal structure, and nothing is known about their behavior under high-pressure conditions. For instance, are they stable under hydrostatic high-pressure? The crystal structure, structural stability, and parameters like the bulk modulus are fundamental descriptors of solids and a necessary input for accurately characterizing other physical properties.

Part of the recent interest in lanthanide monoxides comes from their magnetic properties at low temperatures [6]. For instance, lanthanum monoxide, LaO, has been reported to be a ferromagnetic semiconductor with a narrow band gap of 0.11 eV and a Curie temperature of 130 K. The intense search of high-temperature superconductivity using high-pressure [7] include also compounds like LaO [8]. This compound has been reported to exhibit stronger superconductivity compared to lanthanide monochalcogenides, with a critical temperature of approximately 5 K [8]. The superconductivity of LaO and other lanthanide monoxides can be tuned, to enhance the critical temperature, by the decrease of the unit-cell volume, which can be achieved applying an external pressure [9]. Thus, a characterization of the influence of pressure in the crystal structure is fundamental for improving understanding of high-pressure superconductivity [10].

Beyond their superconducting properties, lanthanide monoxides also attract interest due to their potential applications in various fields including chemistry, biology, medicine, and the high-technology industry [11]. Another reason for interest in lanthanide monoxides is that they can be used as models to simulate the properties of transuranic element monoxides, which have gaining considerable attention [12], but are more difficult to be studied due the radioactivity of transuranic elements.

With the aim of contributing to the understanding of lanthanide monoxides we have performed a systematic computational study considering the monoxides of fifteen lanthanides (from La to Lu). Calculations have been performed using the density-functional theory and the Quantum Espresso open-source package [13]. We have performed calculations under the general-gradient approximation (GGA) [14] and the local-density approximation (LDA) [15] to check the accuracy of both functionals to describe lanthanide monoxides. For modeling the structure, we have evaluated three candidate structures, the cubic B1 (NaCl-type), B2 (CsCl-type), and B3 (ZnS-type) structures, which have been reported as possible structures for lanthanide monoxides [3,6,16,17]. Comparing with experiments, we have found that GGA is the approximation that best describes lanthanide monoxides. By means of total-energy calculations, we have also obtained that for the fifteen studied compounds the most stable structure is B1. However, we have also concluded that all the lanthanide monoxides are expected to undergo a B1-B2 transition under high-pressure conditions. The change of the unit-cell volume for all compounds have been also calculated. From these results, an isothermal equation of state (EOS) has been obtained for each compound. The results will be systematically discussed.

## 2. Computational details

To investigate the structural properties and phase transitions of lanthanide monoxides, we have conducted *ab-initio* calculations at zero temperature using the density-functional theory with the pseudopotentials and plane-wave method (PP-PW) as implemented in the Quantum-Espresso (QE) code. In this approach, the ionic cores have been represented using ultra-soft pseudopotentials from the Standard Solid State Pseudopotentials Library (SSSP). Based on convergence tests, we have established a kinetic energy cutoff of 85 Ry for the plane waves and a cutoff of 1190 Ry cutoff for the charge density and potential. These values ensured reliable results across all computations. The traditional Monkhorst-Pack scheme [18] has been employed by using a dense grid of 10 × 10 × 10 to sample the Brillouin zone. In our total-energy calculations we have used the Generalized Gradient Approximation (GGA) and the Local Density Approximation (LDA) for the exchange-correlation energy for the case of B1 phase. Since we have found that the first method gives more accurate results for the ambient-pressure structure, for studying the stability under pressure of the studied compounds only GGA has been used.

During the generation of pseudopotentials, all but one of the *f* electrons for the lanthanide atoms have been frozen into the core, while for oxygen, the $2s^2$ and $2p^4$ electrons have been explicitly treated as valence electrons. For each monoxide, we have calculated the energy across various lattice constants and corresponding volumes to determine the equilibrium properties of their cubic structures. The energy-volume data points have been fitted using the Birch-Murnaghan equation of state [19], yielding the equilibrium volume, equilibrium energy, bulk modulus, and the pressure derivative of the bulk modulus for each structure. Additionally, we have computed the enthalpy as a function of pressure to determine the occurrence of phase transitions and the transition pressures. To explore the high-pressure stability, we have focused on the three cubic phases reported as stable or metastable phases in the literature [3,6,16,17]: B1 (NaCl-type), B2 (CsCl-type, B2), and B3 (ZnS-type), which are shown in Figure 1. The B1 structure is described by space group $Fm\overline{3}m$ with the lanthanide atom at the 4a Wyckoff position (0,0,0) and the oxygen atoms at the 4b Wyckoff position (1/2,1/2,1/2). The B2 structure is described by space group $Pm\overline{3}m$ with the lanthanide atom at the 1a Wyckoff position (0,0,0) and the oxygen atoms at the 1b Wyckoff position (1/2,1/2,1/2). The B3 structure is described by space group $F\overline{4}3m$ with the lanthanide atom at the 4a Wyckoff position (0,0,0) and the oxygen atoms at the 4c Wyckoff position (1/4,1/4,1/4).

### 3. Results and discussion

Figures 2, 3, 4, and 5 show the calculated enthalpy versus pressure for the fifteen studied compounds using the GGA approximation. In all the figures it can the seen that at 0 GPa, the B1 structure is the one with the lowest enthalpy, indicating that at ambient pressure it is the thermodynamically stable structure in the fifteen studied compounds. This result agrees with the fact that most experiments have reported lanthanide monoxides with this structure. Interestingly, at 0 GPa the structure closest in enthalpy to B1 is B3, which is consistent with the fact that it has been observed as a metastable phase in compounds like GdO, SmO, and EuO [3,17].

The calculated unit-cell parameters for those compounds that have been experimentally reported are summarized in Table 1 and compared with experiments [9,20-34]. In the table we include results from GGA and LDA calculations. In Table 2 we have summarized the results we have obtained for the fifteen studied compounds, including those never studied experimentally, for the interest of future studies. Table 1 shows that GGA agrees better with experiments than LDA, providing a more accurate description of the crystal structure. The LDA accurately reproduces the unit-cell parameter trend of the GGA calculations and experiments, but the magnitudes of the unit-cell parameters are smaller. In Table 1 we also compare our results with previous results obtained from GGA calculations carried out by Shafiq *et al.* using the WIEN2k and the PBE functional [5]. The WIEN2k calculations tend to underestimate the value of the lattice parameter, in special for CeO and PrO. However, they agree better with experiments than our calculations for EuO. The differences between our calculations and those performed with WIEN2k [5] might be related to the use of a different functional. On the other hand, the underestimation of the lattice parameter by the LDA method is common in many systems [35] and apparently is related to the limitations of LDA to describe the exchange-correlation energy in systems with $f$ electrons, like lanthanides. In fact, a similar overestimation of the unit-cell parameter by LDA has been observed for lanthanide metals, for which GGA gives also a more accurate description [36]. Based on this fact, the existence of phase transitions under high-pressure will be discussed using the GGA method.

It should be also noted that EuO is the only compounds, for which the largest difference has been found between our GGA calculations and experiments. In EuO the relative difference between both results is 4.8 %. This might be related to that $Eu^{2+}$ has the most accessible divalent oxidation state because of the half-filled $4f^7$ electronic configuration of Eu and, consequently, a higher stabilization from exchange energy than other lanthanide elements [36]. This stabilization leads to significant anomalies in atomic radii of Eu and its physical properties, including the lattice parameter which is the largest among lanthanides [37]. The differences between DFT calculations and experiments for the lattice parameter of EuO have been for reported for europium chalcogenides and pnictides [38], being a common feature for divalent europium compounds, being considered to the be related to a non-accurate description by theory of $f$-electron delocalization in Eu [39].

We will now discuss the influence of high-pressure in the crystal structure. In Figures 2, 3, 4, and 5 show that the difference of enthalpy between phases B1 and B3 increases under compression, making B3 not a candidate for a HP phase. In contrast, the difference of enthalpy between B1 and B2 decreases under compression, then B2 becoming the most stable structure beyond a critical pressure. The B1-B2 transition is accompanied by an increase in the coordination number in the first coordination sphere of the lanthanide atoms from six to eight (See Fig. 1). The predicted phase transition is consistent with the high-pressure characteristics exhibited by other monoxides such as SrO, CaO, Cd and MgO, which undergoes the B1-B2 transitions at 32. 56, 176, and 198 GPa, respectively [40 41]. The transition pressures are summarized in Table 2. For most of the studied compounds, they range from 71 to 135 GPa, pressures that can be routinely obtained now-a-days in laboratories using diamond-anvil cells. The only two outliers are YbO, for which a transition pressure of 29 GPa is predicted, and LuO for which a transition pressure of 209 GPa is predicted. Our results make YbO the best candidate for the experimental search of the B1-B2 transition, because it happens at pressure quite accessible for the current experimental methods. Notice that YbO has been synthesized by three different research groups, therefore it would be possible to perform high-pressure x-ray diffraction experiments to test or prediction of the B1-B2 transition. Another result to highlight from Table 2 is that the B1-B2 transition involves a large volume collapse, which is typical of this transition, which is a reconstructive first-order transition [42]. Summing up, we consider that out findings highlight the importance of leveraging pressure as an additional dimension in the study of lanthanide monoxides predicting the existence of a B1-B2 transition for all the members of the family and underscore the potential of using first-principles calculations to guide experiments involving materials difficult to synthesize.

Additionally, we have determined the pressure dependence of the unit-cell volume with pressure, which has been fitted for each compound using a third-order Birch-Murnaghan equation of state [19]. From the fitting we obtained the bulk modulus at zero pressure, $B_0$, and its pressure derivative, $B_0'$. The bulk modulus is an important mechanical property to determine the substrate influence on the properties of a thin film like those prepared from lanthanide monoxides. The obtained EOS parameters are summarized in Table 2. All compounds show bulk modulus between 125 and 152 GPa, meaning that lanthanide monoxides are more incompressible than CaO, with $B_0 = 111$ GPa [43], and slightly less incompressible than MgO, with $B_0 = 156$ GPa [44]. Our results show a smooth variation of the bulk modulus when going through the lanthanide series, with the maximum values of the bulk modulus obtained for compounds with a lanthanide atom near the center of the series. The compounds with the smaller bulk modulus are LaO, CeO, YbO, and LuO. Our calculations of the bulk modulus contrast with those obtained through elastic-constant calculations using the cubic-elastic software [5], which are shown for comparison in Table 2. The bulk moduli obtained from such calculations shown an unusual scattering of results with changes in the bulk modulus of up to 110 GPa when moving from one compound to the next one within the family, giving a bulk modulus as low as 90 GPa for EuO and as large as 277 GPa for YbO. It is not reasonable for the last compound to have a bulk modulus that is triple that of the first compound. This fact makes us confident in our reported bulk moduli. Nevertheless, high-pressure x-ray diffraction experiments are needed to confirm our results. We hope this work will motivate them. To conclude, for completeness we present in Table 3, the unit-cell parameters at 0 GPa we have obtained for the B2 and B3 phases as well as the EOS parameters for the third-order Birch-Murnaghan EOS [19] that describes the pressure dependence of the volume for both phases.

## 4. Conclusions

In this study, we have conducted a comprehensive computational analysis of the structural properties and phase transitions of lanthanide monoxides using density-functional theory with both the Generalized Gradient Approximation (GGA) and the Local Density Approximation (LDA). Our results confirm that the B1 (NaCl-type) structure is the most stable phase at ambient pressure for all fifteen studied lanthanide monoxides, which aligns with experimental observations.

We have also found that the GGA method is more accurate in reproducing experimental lattice parameters compared to the LDA method, which systematically underestimates lattice parameters. Under hydrostatic high-pressure conditions, we have predicted that all lanthanide monoxides will undergo a phase transition from the B1 to the B2 (CsCl-type) structure. The transition pressures vary across the series, with YbO showing the lowest transition pressure at 29 GPa, making it an excellent candidate for experimental verification using current high-pressure techniques. The predicted large volume collapse associated with this transition is consistent with the behavior of similar monoxides and underscores the reconstructive nature of the B1-B2 transition. We have also calculated the bulk moduli of the studied compounds, finding that lanthanide monoxides are generally more incompressible than CaO but slightly less incompressible than MgO. The smooth variation of bulk modulus values across the lanthanide series provides further insight into the material properties and potential applications of these compounds.

Overall, our findings enhance the understanding of lanthanide monoxides, highlighting the importance of pressure as a variable in their structural characterization. The study also emphasizes the utility of first-principles calculations in guiding future experimental efforts, particularly in the synthesis and high-pressure analysis of these challenging materials. We hope that our results will stimulate further experimental work to verify the predicted phase transitions and explore the rich physics of lanthanide monoxides under extreme conditions.

**CRediT authorship contribution statement**

Sergio Ferrari: Investigation (equal); Methodology (equal); Writing – review & editing (equal). Daniel Errandonea: Investigation (equal); Methodology (equal); Writing – review & editing (equal).

**Declaration of competing interest**

The authors declare that they have no known competing financial interests or personal relationships that could have appeared to influence the work reported in this paper.

**Data availability**

The data that support the findings of this study are available from the corresponding author upon reasonable request.

**Acknowledgments**

S. Ferrari would like to thank Comisión Nacional de Energía (CNEA) for hiring him as a scientific researcher as supporting his research. D.E. thanks the financial support from the Spanish Ministerio de Ciencia, Innovación y Universidades, MCIU (10.13039/501100011033) under grants RED2022-134388-T and PID2022-138076NB-C41, D. E. also thanks the financial support of Generalitat Valenciana. through grants PROMETEO CIPROM/2021/075-GREENMAT and MFA/2022/007. This study forms part of the Advanced Materials program and is supported by MCIU with funding from the European Union Next Generation EU (PRTR-C17.I1) and by the Generalitat Valenciana.

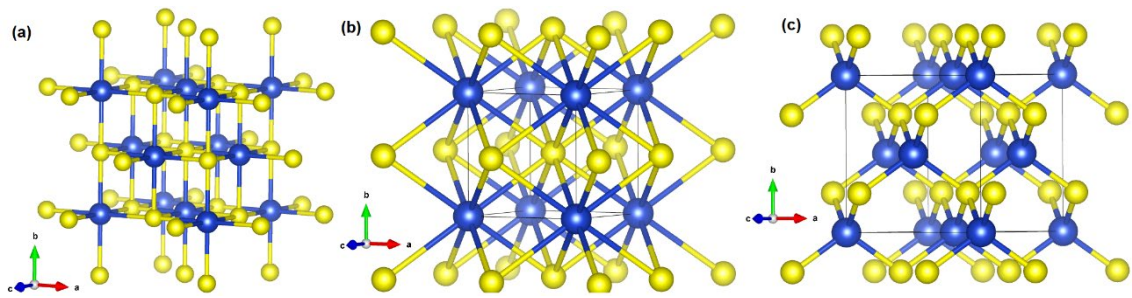

**Figure 1:** Crystal structure of the (a) B1, (b) B2, and (B3) phases. The blue (yellow) spheres represent the lanthanide (oxygen) atoms.

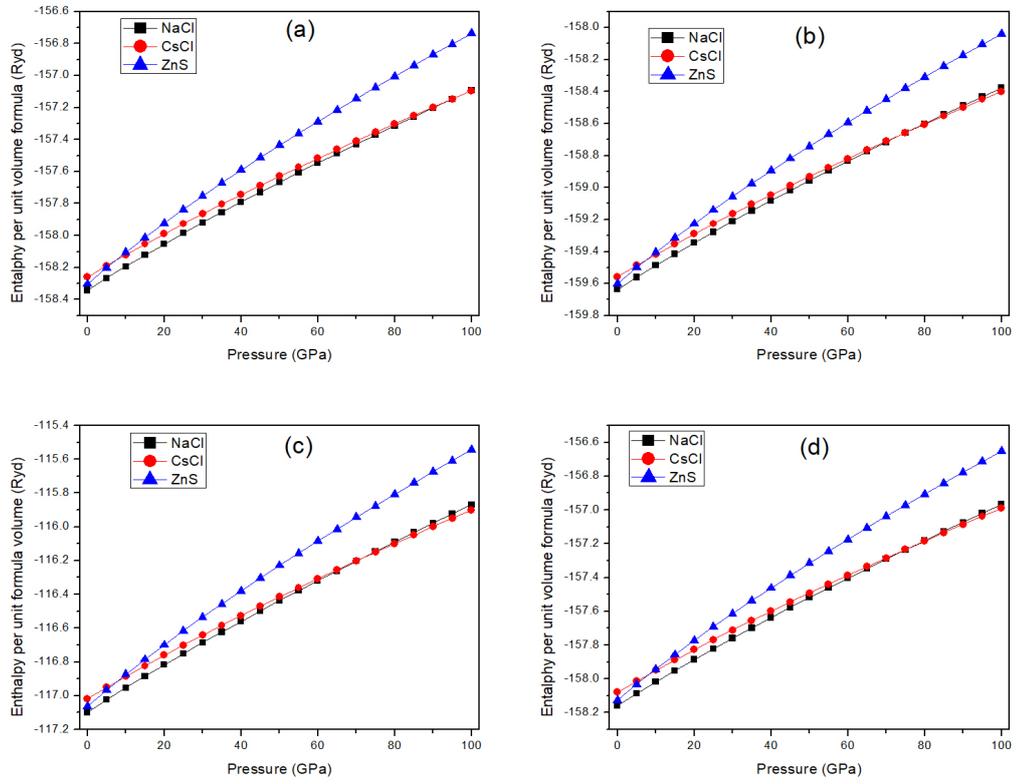

**Figure 2:** Enthalpy versus pressure for (a) LaO, (b) CeO, (c) PrO, and (d) NdO.

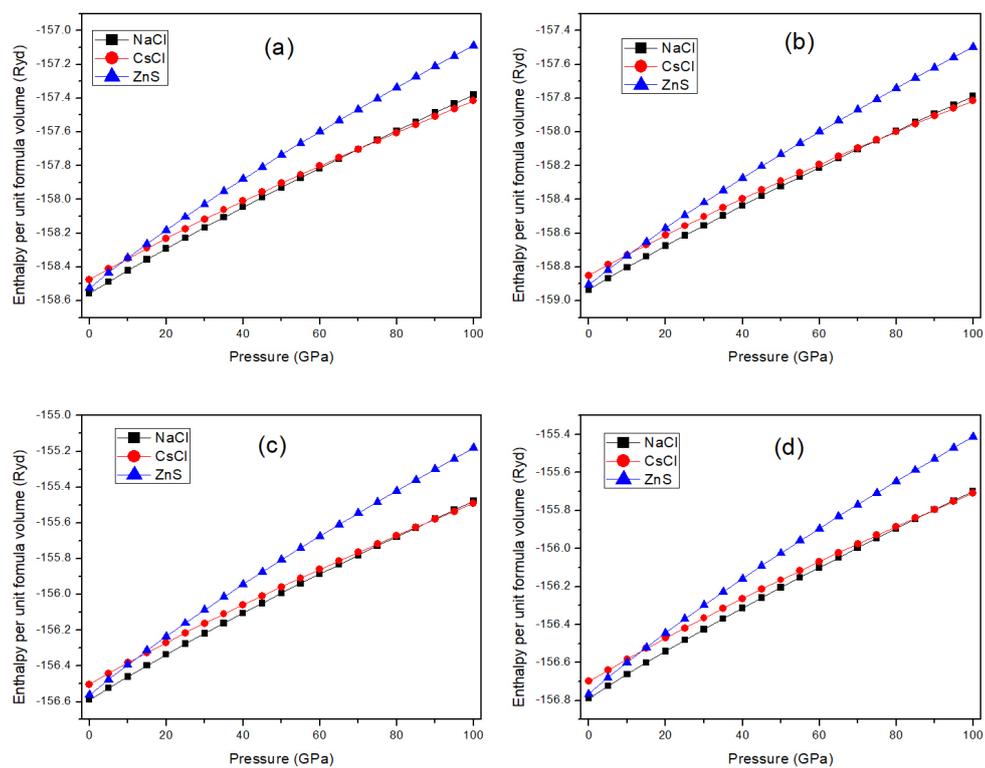

**Figure 3:** Enthalpy versus pressure for (a) PmO, (b) SmO, (c) EuO, and (d) GdO.

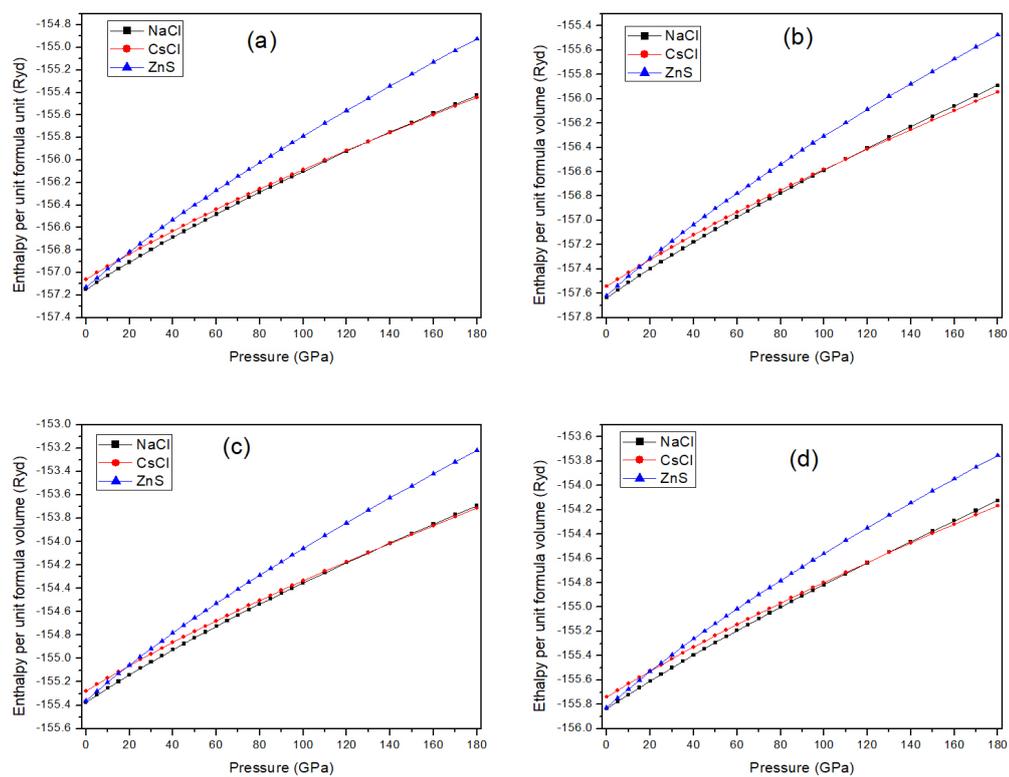

**Figure 4:** Enthalpy versus pressure for (a) TbO, (b) DyO, (c) HoO, and (d) ErO.

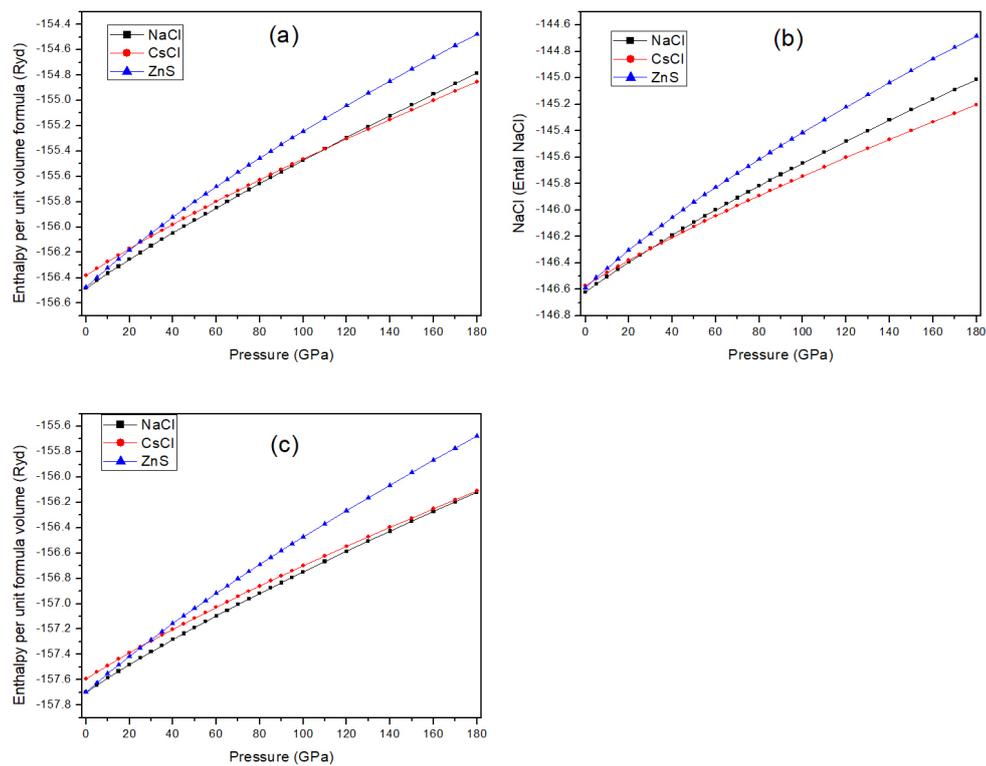

**Figure 5:** Enthalpy versus pressure for (a) TmO, (b) YbO, and (c) LuO.

**Table 1.** Experimental and calculated unit-cell constant (in Å) of experimentally reported lanthanide monoxides in the B1 phase. The references from which the experimental results were extracted are indicated in the table. The average of experimental values is also shown with the standard deviation between brackets. Results from previous calculations [5] obtained from the reported density are also included for comparison.

|  | Experiments | Average of experimental values | QE calculated value using GGA This work | QE calculated value using LDA This work | WIEN2k calculated using GGA Ref. 5 |
|---|---|---|---|---|---|
| LaO | 5.144 [20]<br>5.125 [21]<br>5.198 [9] | 5.156(4) | 5.1643 | 5.0583 | 5.1102 |
| CeO | 5.089 [20]<br>5.089 [22] | 5.089(1) | 5.1312 | 5.0244 | 4.9931 |
| PrO | 5.031 [20]<br>5.031 [22]<br>5.0316 [23] | 5.0312(3) | 5.0677 | 4.9600 | 4.9295 |
| NdO | 4.994 [20]<br>4.9960 [22]<br>5.0101 [24] | 5.000(8) | 5.0144 | 4.9051 | 4.9510 |
| SmO | 4.943 [20]<br>4.9414 [23]<br>4.9414 [24]<br>5,015-5,050 [25]<br>4.9883 [3]<br>4.94 [26] | 4.97(4) | 4.9256 | 4.8141 | 4.9592 |
| EuO | 5.142 [25]<br>5.1439 [3]<br>5.141 [27]<br>5.1419 [28] | 5.1416(6) | 4.8867 | 4.6103 | 4.9954 |
| GdO | 4.99 [29] | 4.99 | 4.8542 | 4.73634 |  |
| TbO | 4.92 [30] | 4.92 | 4.8243 | 4.58015 | 4.8009 |
| HoO | 4.904 [31] | 4.904 | 4.7629 | 4.6367 | 4.7661 |
| YbO | 4.877 [32]<br>4.87 [33]<br>4.88 [34] | 4.876(5) | 4.7205 | 4.5767 | 4.6566 |

**Table 2.** Unit-cell constant ($a$), bulk modulus at zero pressure ($B_0$), pressure derivative ($B_0{}'$), and pressure of B1-B2 phase transition of the fifteen lanthanide monoxides. We also report the relative change of the unit-cell volume per formula unit at the B1-B2 transition defined as $\Delta V = (V_{B2} - V_{B1})/V_{B1}$ All results have been calculated with Quantum Espresso using the GGA method. The bulk moduli from Ref. 5 are included for comparison.

| Lanthanide monoxide | $a$ (Å) | $B_0$ (GPa) | $B_0{}'$ | B1-B2 transition pressure (GPa) | $\Delta V$ at the B1-B2 transition | Bulk modulus Ref. [5] (GPa) |
|---|---|---|---|---|---|---|
| LaO | 5.1643 | 125.1 | 4.51 | 96 | -0.082 | 102.663 |
| CeO | 5.1312 | 128.0 | 4.68 | 75 | -0.076 | 162.659 |
| PrO | 5.0677 | 133.6 | 5.17 | 71 | -0.078 | 146.852 |
| NdO | 5.0144 | 136.9 | 4.95 | 77 | -0.079 | 162.557 |
| PmO | 4.9673 | 140.5 | 6.06 | 71 | -0.078 | - |
| SmO | 4.9256 | 138.5 | 6.14 | 77 | -0.078 | 108.327 |
| EuO | 4.8867 | 139.1 | 5.5 | 86 | -0.077 | 94.287 |
| GdO | 4.8542 | 137.5 | 5.63 | 90 | -0.077 | - |
| TbO | 4.8243 | 142.4 | 3.02 | 133 | -0.077 | 129.013 |
| DyO | 4.7916 | 148.7 | 4.84 | 110 | -0.076 | - |
| HoO | 4.7629 | 151.6 | 3.38 | 135 | -0.075 | 125.880 |
| ErO | 4.7364 | 149.4 | 5.64 | 124 | -0.073 | 106.973 |
| TmO | 4.7070 | 146.6 | 6.35 | 111 | -0.072 | - |
| YbO | 4.7205 | 124.6 | 4.44 | 29 | -0.159 | 226.067 |
| LuO | 4.6566 | 131.1 | 3.01 | 209 | -0.069 | - |

**Table 3.** Calculated values of the unit-cell constant at zero pressure (*a*), bulk modulus at zero pressure ($B_0$), and its pressure derivative ($B_0$') for the B2 and B3 structures of the lanthanide monoxides.

| Lanthanide monoxide | *a* B2 structure (Å) | $B_0$ B2 structure (GPa) | $B_0$' B2 structure | *a* B3 structure (Å) | $B_0$ B3 structure (GPa) | $B_0$' B2 structure |
|---|---|---|---|---|---|---|
| LaO | 3.1614 | 121.0 | 4.60 | 5.6698 | 81.6  | 3.95 |
| CeO | 3.1482 | 123.3 | 4.65 | 5.6300 | 86.2  | 4.24 |
| PrO | 3.1072 | 128.0 | 4.64 | 5.5624 | 89.5  | 4.13 |
| NdO | 3.0737 | 131.7 | 4.69 | 5.5053 | 93.0  | 4.20 |
| PmO | 3.0451 | 134.3 | 4.56 | 5.4553 | 95.7  | 4.09 |
| SmO | 3.0204 | 135.7 | 4.62 | 5.4103 | 98.4  | 4.08 |
| EuO | 2,9969 | 137.4 | 4.57 | 5.3667 | 101.3 | 4.14 |
| GdO | 2.9778 | 139.5 | 4.39 | 5.3290 | 103.7 | 4.02 |
| TbO | 2.9589 | 141.9 | 4.40 | 5.2920 | 106.4 | 4.43 |
| DyO | 2.9403 | 144.7 | 4.61 | 5.2565 | 108.4 | 4.13 |
| HoO | 2.9230 | 147.6 | 5.01 | 5.2229 | 109.5 | 4.60 |
| ErO | 2.9089 | 148.1 | 5.23 | 5.1948 | 110.5 | 4.08 |
| TmO | 2.8924 | 143.2 | 4.85 | 5.1623 | 111.8 | 3.59 |
| YbO | 2.8071 | 133.0 | 4.44 | 5.1267 | 85.9  | 4.10 |
| LuO | 2.8641 | 142.6 | 4.89 | 5.0996 | 120.1 | 4.56 |